# Magnetic interactions and magnetocaloric effect in $(La_{0.5}Pr_{0.5})_{0.6}Ba_{0.4}MnO_3$: Effect of A-site co-doping

Nguyen Thi My Duc,[a,b,c,]* Chang-Ming Hung,[a] Ngo Thu Huong,[b] and Manh-Huong Phan[a,]**

[a] Department of Physics, University of South Florida, Tampa, Florida 33620, USA

[b] Department of Physics, VNU University of Science, 334 Nguyen Trai, Hanoi, Vietnam

[c] The University of Danang, University of Science and Education, 459 Ton Duc Thang, Danang, Vietnam

**Abstract:** We report on the structural, magnetic, and magnetocaloric properties of polycrystalline $(La_{0.5}Pr_{0.5})_{0.6}Ba_{0.4}MnO_3$ (LPBMO), which was fabricated by a conventional solid-state reaction method. LPBMO undergoes a second-order paramagnetic to ferromagnetic (PM-FM) phase transition around a Curie temperature $T_C \sim 277$ K. The maximum magnetic entropy change $-\Delta S_M^{max}$ is $\sim 3.22$ J kg$^{-1}$ K$^{-1}$ for a magnetic field change of 5 T. Based on the modified-Arrott plot and the iterative Kouvel – Fisher methods, a set of critical exponents ($\beta = 0.514 \pm 0.010$, $\gamma = 1.164 \pm 0.013$, and $\delta = 3.265 \pm 0.023$) are determined. These values are close to those expected for the mean field model with long range interactions below $T_C$ and for the 3D-Ising model with short range interactions above $T_C$. Effect of A-site co-doping (La,Pr) on the magnetic and magnetocaloric properties of LPBMO is analyzed and discussed.

*Corresponding author: ntmduc@ued.udn.vn (N.T.M. Duc)

**Corresponding author: phanm@usf.edu (M.H. Phan)

# INTRODUCTION

Magnetic refrigeration (MR), which is a cooling technology based on the magnetocaloric effect (MCE) of a magnetic solid substance as a refrigerant, has attracted growing attention as it provides a promising solution to conventional gas compression (CGC) based refrigerators. The MR technology possesses notable advantages, such as a higher cooling efficiency, compactness, and environmental affability [1-8]. There are three important figures-of-merit in selecting magnetocaloric materials as magnetic refrigerants; the adiabatic temperature change ($\Delta T_{ad}$), the magnetic entropy change ($\Delta S_M$), and the refrigerant capacity (*RC*). Upon the application of a dc magnetic field, $\Delta T_{ad}$ or $\Delta S_M$ should be as large as possible, which typically peaks around the magnetic phase transition temperature of a magnetic material. While $\Delta T_{ad}$ is not often measured experimentally, $\Delta S_M$ calculated through Maxwell equation from *M-H* isotherms provides a quick screening of the usefulness of a magnetocaloric material. In addition to evaluating the $\Delta S_M$, the widely used parameter is *RC*, which represents an amount of heat transferred between the cold sink and hot sink in an ideal refrigerant cycle [9-10]. The *RC* depends on the magnitude of $\Delta S_M$ and its temperature dependence, which are governed by types of magnetic phase transition. An example of this is $Pr_{0.5}Sr_{0.5}MnO_3$ [11], which exhibits two different temperature ranges; the first order magnetic transition (FOMT) at a low temperature ($T_N \sim 150$ K) and the second order magnetic transition (SOMT) at a high temperature ($T_C \sim 250$ K). For the case of FOMT, a large magnetic entropy change is achieved but it is limited to a narrow temperature region. By contrast, despite having a smaller magnetic entropy change, the SOMT displays a broader $\Delta S_M(T)$ over a wider temperature region, resulting in a larger *RC*. For $Pr_{0.5}Sr_{0.5}MnO_3$, the *RC* values are ~145 J kg$^{-1}$ and ~200 J kg$^{-1}$ for FOMT and SOMT, respectively [11].

Mixed-valence perovskite rare-earth manganese oxides of $RE_{1-x}M_x MnO_3$ ($RE$ is a trivalent rare-earth cation, e.g. $RE^{3+}$ = $La^{3+}$, $Nd^{3+}$, $Pr^{3+}$... and M is a divalent alkali cation, e.g. $M^{2+}$ = $Ba^{2+}$, $Sr^{2+}$, $Pb^{2+}$...) have been extensively studied, due to their tunable MCEs, ease of preparation, and cost-effectiveness [10,12-31]. In pure manganites ($LaMnO_3$, $PrMnO_3$), the transition metal atom is predominantly in the state $Mn^{3+}$ [32-34]. The substitution of divalent alkali $M^{2+}$ ions for $RE^{3+}$ ions produces holes in the $e_g$ band which leads to coexistence of $Mn^{3+}$ and $Mn^{4+}$ ions. It has been experimentally shown that while pure $REMnO_3$ is an insulating antiferromagnet, a hole-doped perovskite manganite $RE_{1-x}M_xMnO_3$ can be converted into a conductive ferromagnet [35]. This has been explained using the double-exchange interaction (DE) theory [36,37]. Based on this DE theory, the transfer of an itinerant $e_g$ electron between the neighboring Mn sites (local $t_{2g}$ spins) through the $O^{2-}$ anion results in a ferromagnetic interaction due to the on-site Hund's rule coupling. When placing the material into an external magnetic field, the magnetic field will force the local $t_{2g}$ spins to align, thus enhancing the ferromagnetic phase. Other theories have also been proposed to interpret the structure-influenced magnetic behavior in $RE_{1-x}M_xMnO_3$, such as Jahn-Teller effect [38], antiferromagnetic super-exchange (SE), orbital and charge ordering [39]. Thus, the magnetic and magnetocaloric behaviors in $RE_{1-x}M_xMnO_3$ depend strongly on $Mn^{3+}/Mn^{4+}$ ratio [40-43].

Hole-doped manganese oxides, including $Pr_{1-x}Ba_xMnO_3$ [16-20] and $La_{1-x}Ba_xMnO_3$ [19-24], have shown the large magnetocaloric effects. While $La_{1-x}Ba_xMnO_3$ display high $T_C$ ($Tc \sim$ 292 – 342 K) [19-24], the $T_C$ values of $Pr_{1-x}Ba_xMnO_3$ are considerably smaller ($Tc \sim$ 130 – 235 K) [16-20]. However, the former gives a low $RCP$ value (~81.7 J kg$^{-1}$ at 1 T for $La_{0.6}Ba_{0.4}MnO_3$) [16], while the $RCP$ value for the later is significantly lower (~ 30 J kg$^{-1}$ at 1 T for $Pr_{0.6}Ba_{0.4}MnO_3$) [23]. There have also been cases of co-doping two elements into the $M^{2+}$ position such as in $La_{0.7}Ca_{0.3-x}Sr_xMnO_3$ [25] or combination of two elements in the $RE^{3+}$ position such as in $La_{1-x}Pr_xMnO_3$ [26,27]. However, a clear understanding of *A*-site co-doping effects on the magnetic and magnetocaloric properties, including correlation between the magnetic ordering/ interaction determined by a set of critical exponents and the MCE around the magnetic phase transition temperature, in these systems has been not reached.

To gain deeper insight into this, two $RE^{3+}$ components ($La^{3+}$ and $Pr^{3+}$) were combined with a 50-50 ratio, and the *A*-site co-doping effects of $(La_{0.5}Pr_{0.5})_{0.6}Ba_{0.4}$ on the magnetic and magnetocaloric properties of $(La_{0.5}Pr_{0.5})_{0.6}Ba_{0.4}MnO_3$ have been investigated systematically. We have found a large magnetic entropy change at $T_C$ near room temperature, with a high relative cooling power, indicating that this material is a prospective candidate for magnetic refrigeration. An analysis of the correlation between the magnetic ordering/ interaction determined by critical exponents and the MCE around the magnetic phase transition has been performed.

## EXPERIMENT

The $(La_{0.5}Pr_{0.5})_{0.6}Ba_{0.4}MnO_3$ (LPBMO) sample was prepared by a conventional solid-state reaction method from a stoichiometric mixture of high purity oxides $La_2O_3$, $Pr_2O_3$, $Ba_2O_3$, $MnO_2$ powders up to 99.9%. The sample was pre-sintered at 1000°C for 10h. The heated sample was cooled to room temperature, reground to fine particles, pressed into pallets and finally sintered at 1250°C for 10h.

The structure of this sample was examined in a Bruker D5005 X-ray diffractometer using $Cu/K_\alpha$ radiation. The microstructure and chemical composition were studied on a scanning electron microscope (SEM) equipment–450–FEI. Magnetic measurements including hysteresis loops and isothermal magnetization curves of the sample were performed in a vibrating sample magnetometer (VSM) DSM-880 in magnetic fields up to 1.35 T. The temperature dependent magnetization $M(T)$ curves were measured on a SQUID device at temperature range from 5 K to 350 K. The magnetization measurements were measured by means of the vibrating sample magnetometer option of a Quantum Design commercial physical property measurement system (PPMS) in magnetic fields of up to 5 T and the temperature was varied from 100 K to 319 K. The measurement step was 1 K around the phase transition (274 K – 285 K) and an increment of 3 K in the remaining region (100 K – 274 K and 285 K – 319 K). The molecular weight of the $(La_{0.5}Pr_{0.5})_{0.6}Ba_{0.4}MnO_3$ sample is 241.81 g/mol.

## RESULTS AND DISCUSSION

**Structural properties**

The structural properties of LPBMO have been characterized by XRD and SEM. The results are displayed in **Fig. 1** and its inset. **Fig. 1** shows the room-temperature XRD pattern taken from 10° to 80° ($2\theta$) with a scanning rate of 0.9°/min. The XRD pattern reveals the single phase nature of the sample without a detectable impurity peak. All the diffraction peaks are indexed by a cubic perovskite structure with the lattice constant $a$ = 3.93 Å, which is fully consistent with that reported previously [21,28,29]. LPBMO exhibits a polycrystalline behavior with the maximum intensity for the (110) reflection. We have also calculated the average crystallite size of LPBMO based on the Debye – Scherrer equation: $D = \frac{0.9\lambda}{\beta\cos\theta}$, where $\lambda = \lambda_{Cu-K\alpha}$ = 1.54 Å is the wave-length of X-ray using Cu-K$_\alpha$ radiation anode, $d$ is the full width at half maximum intensity of the peak and $\theta$ is the diffraction angle. The calculated average crystallite size is 93.4 nm. The inset of **Fig. 1** displays a SEM image of the surface of LPBMO, indicating that the grains are quite homogeneous with the crystalline size of ~100 nm, which is also close to that determined from the XRD pattern.

The Jahn-Teller effect (JT) occurs on metal ions that contain an odd number of electrons in the $e_g$ level, e.g. $Mn^{3+}$ ions in an octahedral crystal field with $3d^4$ orbital ($t_{2g}^3 e_g^1$) [44]. The JT effect causes an ideal cubic structure to be distorted into the orthorhombic structure. However, $Mn^{4+}$ ion has only three electrons localized on $t_{2g}$, so it is not affected by the JT effect. It has been reported that $La_{2/3}Ba_{1/3}MnO_3$ [22], $La_{0.7}Ba_{0.3}MnO_3$ [28], and $La_{0.5}Ba_{0.5}MnO_3$ [29] $O_3$ have a cubic structure, so negligible crystal lattice distortion is expected and $Mn^{4+}$ dominates in this case. By contrast, $Pr_{1-x}Ba_xMnO_3$ has an orthorhombic structure [16-19], so $Mn^{3+}$ dominates in this case. It also depends on the so-called tolerance factor *t*. The substitution of $La^{3+}$ ions by $Pr^{3+}$ leads to a reduction of the tolerance factor due to the decrease in the mean ionic ratio. That will reduce geometric stability of $Pr_{1-x}Ba_xMnO_3$. Thus, JT distortion usually appears in the orthorhombic structure but is not allowed in the cubic structure that has a higher symmetry. As LPBMO has a cubic structure, JT lattice distortion is much less significant. This suggests that the DE theory can be used to explain the magnetic properties and critical behavior of the present sample.

**Magnetic and magnetocaloric properties**

Fig. 2a (the blue curve on the left hand) presents the temperature dependence of the magnetization, $M(T)$, for LPBMO measured while cooling under an applied magnetic field of $\mu_0 H = 0.05$ T. As can be seen in Fig. 2a, the $M(T)$ curve displays a broaden paramagnetic to ferromagnetic (PM-FM) phase transition around the Curie temperature $T_C \sim 276$ K. The inset of Fig. 2a shows how we have calculated the $T_C$ by taking the minimum of the derivative $dM/dT$. This $T_C$ value is significantly larger than that of $La_{0.5}Pr_{0.5}MnO_3$ (LPMO) ($T_C = 137$ K [26] and $T_C = 245$ K [27]) without $Ba^{2+}$ doping). This shows the influence of $Ba^{2+}$ doping on the rare earth (RE = La, Pr) site. The $T_C$ values of samples containing only one RE element (La or Pr) with $Ba^{2+}$ doping ($Pr_{1-x}Ba_xMnO_3$ – PBMO [16-21] or $La_{1-x}Ba_xMnO_3$ – LBMO [20-24]) were also found to be different from that of $(La_{0.5}Pr_{0.5})_{0.6}Ba_{0.4}MnO_3$. The $T_C$ of LPBMO is significantly larger than that of PBMO ($Tc \sim 130 - 235$ K) [16-20], but smaller than that of LBMO ($Tc \sim 292 - 342$ K) [19-24]. The $T_C$ of $La_{0.5}Pr_{0.5}MnO_3$ with no $Ba^{2+}$ doping is also quite low ($\sim 137$ K [26]), due to the coexistence of both DE and SE interactions, respectively. When the DE interaction dominates over the SE interaction, $T_C$ will be shifted towards a higher temperature range.

The magnetic field dependence of magnetization (the *M-H* loop) for LPBMO measured at room temperature ($T_{room} = 300$ K), as shown in **Fig. 2b**, indicates the room temperature ferromagnetic characteristic of the material. In the low magnetic field regime ($\mu_0H < 150$ mT), the hysteresis curve appears quite clearly, while in the high magnetic field regime ($\mu_0H > 150$ mT), the magnetization increases rapidly as the magnetic field increases. The maximum magnetization at $\mu_0H = 12$ kOe is $M_{max} = 13.37$ A m$^2$ kg$^{-1}$. The coercive field is calculated as $H_C = 3.64$ Oe at room temperature, which is small and therefore beneficial for use as an active magnetocaloric material. In pure manganites ($LaMnO_3$, $PrMnO_3$), the transition metal atom Mn is predominantly in the $Mn^{3+}$ state, so the materials order antiferromagnetically below $T_N$ [32-34]. Since only $Mn^{3+}$ ions are present in these manganites, the interaction between them is primarily explained by the SE theory. The SE interaction may be ferromagnetic or antiferromagnetic, but the DE interaction is only ferromagnetic [36,37]. In this work, LPBMO has a ferromagnetic characteristic at room temperature as seen in **Fig. 2b**. This is because, in doped manganites, DE interactions appear to occur when replacing a $RE^{3+}$ position by divalent ions like $Ba^{2+}$. Because substitution of a divalent alkali for a trivalent $RE^{3+}$ cation will reduce the total charge to retain the neutral charge condition, a part of $Mn^{3+}$ changes into $Mn^{4+}$. With the appearance of $Mn^{4+}$, the electrical conductivity increases and ferromagnetism emerges via the DE mechanism. The results obtained from our study indicate that the $Ba^{2+}$-doping has a great influence on the magnetic property of LPBMO.

The inverse of the magnetic susceptibility, $\chi^{-1}(T) = \mu_0 H/M$, defined from the $M(T)$ curve in the paramagnetic region is also shown in **Fig. 2a** (the red curve on the right hand). Because of the broaden PM-FM phase transition, the $\chi^{-1}(T)$ is only linear at high temperature region above 300 K. According to the Curie-Weiss law in the paramagnetic region: $\chi = \frac{C}{T-\theta}$ with $C$ is the Curie constant specified by $C = \frac{N_A \mu_B^2}{3k_B}\mu_{eff}^2$, where $N_A = 6.022 \times 10^{23}$ mol$^{-1}$ is Avogadro's number, $\mu_B = 9.274 \times 10^{-21}$ emu is the Bohr magneton, and $k_B = 1.38016 \times 10^{-16}$ erg/K (in the CGS system of units) is Boltzmann constant, the Curie-Weiss temperature $\theta$ and the effective magnetic moment $\mu_{eff}$ can be specified. Fitting the linear region of $\chi^{-1}(T)$ yields $\theta = 311$ K and $C = 0.516$ emu K mol$^{-1}$. $\theta = 311$ K is a positive value which confirms the PM-FM phase transition. The broaden PM-FM phase transition is the cause of the difference between $\theta$ and $T_C$. The effective magnetic moment $\mu_{eff}$ can be calculated based on relationship between $C$ and $\mu_{eff}$: $\mu_{eff} = \left(\frac{3k_B C}{N_A}\right)^{1/2} = \sqrt{8C}\mu_B$. From this equation, $\mu_{eff} = 2.03$ $\mu_B$ for LPBMO. For $La_{0.5}Pr_{0.5}MnO_3$ (LPMO), $\mu_{eff} = 3.3$ $\mu_B$ [26]. The $\mu_{eff}$ value of LPBMO is significantly smaller than that of LPMO. The $\mu_{eff}$ value of LPBMO decreases due to the increase of the $Mn^{4+}/Mn^{3+}$ ratio, which elaborates well with the dominant ferromagnetism in this sample. This also proves that the DE interaction is essential, so the crystal structure is highly symmetric (the cubic structure in **Fig. 1**) and the resulting high $T_C$ near room temperature.

To further understand the nature of the PM-FM phase transition, a set of isothermal magnetization $M(\mu_0 H)$ curves of LPBMO was taken at different temperatures around $T_C$ from 100 to 319 K, with a temperature step interval between subsequent isotherms of $\delta T_1 = 1$ K from 274 – 285 K and $\delta T_2 = 3$ K for other temperature ranges, under $\mu_0 H = 0 - 5$ T, as shown in **Fig. 3a**. The sweeping rate of the applied magnetic field was slow enough to ensure that the magnetization curves were obtained in an isothermal process. This set of isothermal magnetization curves is an important measurement from which the magnetic and magnetocaloric properties of materials can be assessed. As can be seen in **Fig. 3a**, the magnetization (*M*) has a large change in magnitude around $T_C$. At the lowest temperature (100 K), even at low magnetic field (~ 0.1 T), *M* increases sharply and tends to saturate at $\mu_0 H > 1.5$ T.

**Figure 3b** presents the *M*(*T*) curves at different applied magnetic fields up to 5 T. The PM-FM phase transition becomes broadened with increase of the applied magnetic field, which is a typical behavior for SOMT materials. From **Fig. 3b,** one can see that the magnetization has a large change in the temperature region 250 - 300 K, where $T_C$ has been found (~ 276 K) from the minimum of the derivative d*M*/d*T*. This leads to an expectation for the maximum entropy change $\Delta S_M^{max}$ to be around this temperature region. In order to have a more visual view of the *M*(*T*) broadening, the filled 2D contour plot of the temperature and applied magnetic field dependences of magnetization is shown in **Fig. 3c**. The magnetic phase transition extends over a wide range temperature 300 – 250 K from the magnetically disordered PM state to the ordered FM state and the sharp change in *M* occurs strongly at ~ 275 K. This is also alternatively viewed from a filled 2D contour plot of d*M*/d*T* (*T*,*H*) in **Fig. 3d**.

The magnetic entropy change $\Delta S_M$ can be specified from the isothermal *M*($\mu_0 H$) curves using the thermal-dynamic Maxwell equation [1]: $\Delta S_M(T, \mu_o H) = \mu_0 \int_0^{H_{max}} \left(\frac{\partial M}{\partial T}\right)_H dH$, where *M* is the magnetization, $\mu_o H$ is the applied magnetic field, and *T* is the temperature, by integrating over the magnetic field. **Fig. 4a** describes the temperature dependence of the magnetic entropy change $-\Delta S_M$ at difference field changes, $\mu_o\Delta H = 0.1 - 5$ T. From **Fig. 4a**, one can clearly see that the $-\Delta S_M(\mu_o H,T)$ curves across the PM-FM phase transition are broad. This well elaborates with the broadened PM-FM transition as seen in the $M(T)$ curves for high fields, 5 T (see **Fig. 3b**). All the $-\Delta S_M(\mu_o H,T)$ curves expose that the maximum value of the magnetic entropy change $\Delta S_M^{max}$ increases with increasing magnetic field. It is also observed that all the peaks appear to occur at almost the same temperature, which is close to the $T_C$, ~ 276 K. For $\mu_o\Delta H = 5$ T, the $-\Delta S_M^{max}$ is found to be 3.22 J kg$^{-1}$ K$^{-1}$ at 276 K. This $-\Delta S_M^{max}$ value (for $\mu_o\Delta H = 1$ T, $-\Delta S_M^{max}$ ~ 0.9 J kg$^{-1}$ K$^{-1}$) is smaller than that of Pr$_{0.6}$Ba$_{0.4}$MnO$_3$ (for $\mu_o\Delta H = 1$ T, $-\Delta S_M^{max}$ ~ 1.5 J kg$^{-1}$ K$^{-1}$) [16] but significantly higher than that of La$_{0.6}$Ba$_{0.4}$MnO$_3$ (for $\mu_o\Delta H = 1$ T, $-\Delta S_M^{max}$ ~ 0.6 J kg$^{-1}$ K$^{-1}$) [23]. This testifies our initial hypothesis that when combining La$^{3+}$ and Pr$^{3+}$, $-\Delta S_M^{max}$ increases significantly compared to La$_{0.6}$Ba$_{0.4}$MnO$_3$, while retaining the $T_C$ near room temperature. The $\mu_o\Delta H$ dependence of $-\Delta S_M^{max}$ is quite linear, as described in **Fig. 4b**, indicating the larger magnetic field entropy changes for the higher magnetic field changes. A large $-\Delta S_M^{max}$ value is one of the parameters resulting in a large relative cooling power *RCP*, which is an important figure-of-merit for calculating the cooling efficiency.

The *RCP* can be defined as Wood and Potter's method: $RCP = -\Delta S_M^{max} \cdot \delta T_{FWHM}$ [45], where $\delta T_{FWHM} = T_{hot} - T_{cold}$ is the temperature difference at the full width at half maximum of the magnetic entropy change curve. A combination of the large values of $-\Delta S_M^{max}$ and $\delta T_{FWHM}$ is expected to result in the large *RCP* value. **Fig. 4c** describes the magnetic field dependence of $\delta T_{FWHM}$. From **Fig. 4c**, it is easy to see that $\delta T_{FWHM}$ increases significantly upon the increasing of $\mu_o\Delta H$, e.g. for $\mu_o\Delta H = 1$ T, $\delta T_{FWHM} = 59$ K and $\mu_o\Delta H = 5$ T, $\delta T_{FWHM} = 84$ K. The *RCP* as a function of $\mu_o\Delta H$ is performed in **Fig. 4d**. When the applied magnetic field increases, the *RCP* increases significantly, associated with the increase in $-\Delta S_M^{max}(\mu_o\Delta H)$ as can be seen in **Fig. 4b**.

For $\mu_o\Delta H = 5$ T, the *RCP* of LPBMO is 270 J kg$^{-1}$, which is significantly larger than those of Ba$^{2+}$ doped samples containing only *RE*$^{3+}$ [16-24]. Although the $-\Delta S_M^{max}$ of LPBMO is not as large as that of PBMO, the $\delta T_{FWHM}$ is larger, due to the broader magnetic phase transition, resulting in the larger *RCP* of LPBMO. To put the magnetocaloric parameters of the present material and other candidates in comparison, **Tab. I** summarizes the Curie temperature $T_C$, the magnetic field change $\mu_0\Delta H$, the maximum magnetic entropy change $|\Delta S_M^{max}|$, and the relative cooling power *RCP* of these samples.

With the aim of elucidating the critical phenomenon that occurs around the PM-FM phase transition and relating it to the observed $-\Delta S_M(\mu_o H)$ behavior, we have studied the critical exponents near the PM-FM phase transition, as shown below.

**Critical exponents**

In SOMT materials, the critical behavior of the PM-FM phase transition can be defined by a set of three critical exponents, $\beta$, $\gamma$, and $\delta$ [46], which has been correlated with the magnetocaloric behavior of SOMT materials. The first critical exponents $\beta$ is associated with the spontaneous magnetization $M_S(T)$, the second critical exponents $\gamma$ is associated with the initial inverse susceptibility $\chi_0^{-1}(T)$, and the last one $\delta$ is associated with the field dependence of the magnetization of the critical isotherm $M(\mu_o H)$ at $T_C$ [46]. These three exponents can be determined according to the power-law relations [47]:

$$M_S(T) = M_0(-\varepsilon)^\beta \qquad \varepsilon < 0 \quad T < T_C \qquad (1)$$

$$\chi_0^{-1}(T) = \left(\frac{H_0}{M_0}\right)\varepsilon^\gamma \qquad \varepsilon > 0 \quad T > T_C \qquad (2)$$

$$H = DM^\delta \qquad \varepsilon = 0 \quad T = T_C \qquad (3)$$

where $\varepsilon$ is the reduced temperature, $\varepsilon = (T - T_C)/T_C$, and $M_0$, $H_0$, $D$ are the critical amplitudes.

Based on the Arrott – Noakes equation of state $\left(\frac{H}{M}\right)^{\frac{1}{\gamma}} = A\varepsilon + BM^{\frac{1}{\beta}}$ [48], where A and B are material dependent parameters, the Curie temperature $T_C$ can be recalculated using modified Arrott plot (MAP) method. For this purpose, the isothermal magnetization $M(\mu_o H)$ data is reformulated in to $M^{1/\beta}$ [$(\mu_0 H/M)^{1/\gamma}$]. The correct exponents are those that linearize $M^{1/\beta}$ versus $(\mu_0 H/M)^{1/\gamma}$ and $T_C$ is defined from the critical isotherm which passed through the origin.

With the purpose of establishing the corresponding MAP for studying the critical behavior near the PM-FM phase transition, there are 4 theoretical models with the exponential exponent come into consideration: the mean-field model ($\beta = 0.5$, $\gamma = 1.0$) [49], the 3D Ising model ($\beta = 0.325$, $\gamma = 1.240$) [49], the 3D Heisenberg model ($\beta = 0.365$, $\gamma = 1.336$) [49], and the tricritical mean field model ($\beta = 0.25$, $\gamma = 1$) [50]. The spontaneous magnetization $M_S(T)$ and the initial inverse susceptibility $\chi_0^{-1}(T)$ were specified from the intercepts of the linear extrapolation of $M^{1/\beta}$ and $(\mu_0 H/M)^{1/\gamma}$ at the high field isotherms of the modified-Arrott plots, respectively. The critical exponents $\beta$ and $\gamma$ are extracted by fitting $M_S(T)$ data with the relation $M_S \propto (-\varepsilon)^\beta$ in equation (1) and $\chi_0^{-1}(T)$ data with the relation $\chi_0^{-1} \propto \varepsilon^\gamma$ in equation (2), respectively. Using the new $\beta$ and $\gamma$ values this process is repeated until the isotherm that passes through the origin gives $T = T_C$ and the critical exponent $\beta$ and $\gamma$ values reach the stable values. The final modified-Arrott plot in **Fig. 5a** shows that the isotherms achieve good linearity with $\beta_{MAP} = 0.514 \pm 0.010$, $\gamma_{MAP} = 1.164 \pm 0.013$ and $T_{C\text{-MAP}} = 277$ K. Based on the statistical theory, these critical exponents $\beta$, $\gamma$ and $\delta$ must satisfy the Widom scaling relation: $\delta = 1 + \frac{\gamma}{\beta}$ [51]. From the Widom scaling equation, the critical exponent $\delta$ can be calculated, $\delta = 3.265 \pm 0.023$. So, the critical exponents $\beta$, $\gamma$ and $\delta$ determined by MAP method are: $\beta_{MAP} = 0.514 \pm 0.010$, $\gamma_{MAP} = 1.164 \pm 0.013$, and $\delta = 3.265 \pm 0.023$. The final values of $M_S(T)$, $\chi_0^{-1}(T)$ and the fitting curves of them are displayed in **Fig. 5b**.

Furthermore, the critical exponents $\beta$, $\gamma$ and $\delta$ can also be calculated by the Kouvel – Fisher (KF) method by reformulating equations (1) and (2) [52]:

$$M_S(T)\left[\frac{dM_S(T)}{dT}\right]^{-1} = \frac{T - T_C}{\beta} \quad (4)$$

$$\chi_0^{-1}(T)\left[\frac{d\chi_0^{-1}(T)}{dT}\right]^{-1} = \frac{T - T_C}{\gamma} \quad (5)$$

After plotting the $M_S$ vs. $(dM/dT)^{-1}$ and $\chi_0^{-1}$ vs. $d\chi_0^{-1}/dT$, two lines of linear fitting are displayed in **Fig. 5c**. For temperatures below $T_C$, the result for critical exponent $\beta$ is $\beta_{KF} = 0.514 \pm 0.007$ and the Curie temperature is $T_{C1} = 276.56 \pm 2.40$. For temperatures above $T_C$, the result for the critical exponent $\gamma$ is $\gamma_{KF} = 1.164 \pm 0.005$ and the Curie temperature is $T_{C2} = 276.68 \pm 2.52$. The critical exponent $\delta$ can be calculated using Widom scaling equation [51], $\delta_{KF} = 3.265 \pm 0.012$. The critical exponents of LPBMO calculated using the MAP and KF methods are listed in **Tab. II** for a comparison purpose. A good agreement between these two methods confirms that the calculated values of the critical exponents are reliable.

The iterative MAP and KF methods allow an exact determination of a set of true critical exponents. From these two methods, the critical behavior for LPBMO is found with the critical exponent $\beta = 0.514$, which is very close to that of the mean field model ($\beta = 0.5$), and the critical exponent $\gamma = 1.164$ lies between the mean field model ($\gamma = 1$) and the 3D-Ising model ($\gamma = 1.241$). It proves that below $T_C$, it represents a long-range ferromagnetic interaction, belonging to the mean field model and above $T_C$, it describes ferromagnetic short-range interactions belonging to the 3D-Ising model. This appears to be a common feature for doped manganite systems [53-59]. It is generally accepted that the presence of ferromagnetic clusters with a short range interaction in the paramagnetic region in vicinity of the PM-FM phase transition broadens the phase transition and hence the temperature range where the $\Delta S_M(T)$ peak occurs [60,61]. The value of $\Delta S_M$ is usually lower compared to the case with absence of ferromagnetic clusters. Control over density and size of magnetic clusters may provide a plausible approach for designing magnetocaloric materials with enhanced cooling efficiency for active magnetic refrigeration.

Based on the power-law relations at $T = T_C$ (3): $H = DM^\delta$ [47] with ($\varepsilon = 0, T = T_C$), the critical exponent $\delta$ can also be calculated by taking two-sided natural logarithms of equation (3). After that, a log-log plot of applied magnetic field dependence of magnetization $\ln M(\ln(\mu_0 H))$ at temperatures in the vicinity of $T_C$ is shown in **Fig. 5d**. Based on equation (3), the critical exponent $\delta$ can be defined from the inverse slope of the critical isotherm analysis (CIA). A $\ln M(\ln(\mu_0 H))$ linear fitting of the $T_C = 277$ K isotherm yields $\delta_{CIA} = 3.30 \pm 0.05$ for LPBMO. This $\delta_{CIA}$ value is close to that determined from the Widom relation $\delta = 1 + \frac{\gamma}{\beta}$ [51] based on the results of the MAP method ($\delta_{MAP} = 3.265 \pm 0.023$) and the KF method ($\delta_{KF} = 3.265 \pm 0.012$), as summarized in **Tab. II**. The critical exponents calculated from the different methods are almost identical.

For SOMT materials, a scaling law shows the applied magnetic field $\mu_0 H$ dependence of the maximum entropy change $\Delta S_M^{max}$ [62,63]:

$$\Delta S_M^{max} \propto \mu_0 H^n \qquad (6)$$

where $n$ is an addition scaling exponent for the magnetic field dependence of the peak in the magnetic entropy change.

To determine the scaling exponent $n$ using equation (6), the $-\Delta S_M^{max}$ and $\mu_0 H$ values are rescaled to $\ln(-\Delta S_M^{max})$ and $\ln(\mu_0 H)$ by taking two-sided natural logarithms equation (6). The slope of linear fitting of $\ln(-\Delta S_M^{max})$ vs. $\ln(\mu_0 H)$ yields $n = 0.756 \pm 0.029$ for $\mu_0 H = 1.68 - 5$ T. Rescaling the $\mu_0 H$ axis to produce a plot of $-\Delta S_M^{max}$ vs. $(\mu_0 H)^n$ with $n = 0.756$ displays a linear relationship as expected (see **Fig. 6a**).

A relationship between the scaling exponent $n$ and the magnetization exponent $\beta$ and susceptibility exponent $\gamma$ can be expressed as equation: $n = 1 + \frac{\beta - 1}{\beta + \gamma}$ [64]. From these critical exponents $\beta$ and $\gamma$ calculated from MAP and KF methods, $n_{MAP}$ and $n_{KF}$ are defined by using above equation, yields $n_{MAP} = 0.710 \pm 0.033$ and $n_{KF} = 0.710 \pm 0.019$. Besides, we can determine $\beta$ and $\gamma$ from $n$ and the results are 0.561 and 1.240 for $\beta$ and $\gamma$, respectively. The calculated critical exponents are listed in **Tab. II** for ease of comparison. We have found once again that the critical exponent $n$ is close to that of the mean field model.

In addition to the Banerjee's criterion [65] based Arrott plots for determining the type of magnetic phase transition, another method based on the universal curves of $-\Delta S_M(T)$ has been proposed by Franco *et al.* [66]. For SOMT materials, a universal curve can be built up to depict $-\Delta S_M(T)$ at different magnetic fields, $\mu_o \Delta H$. Then, all the $-\Delta S_M(T)$ curves at different values of $\mu_0 \Delta H$ should be collapsed onto a single universal curve, when $\Delta S_M$ is normalized to $\Delta S_M^{max}$ and the temperature axis needs to be rescaled as [66]:

$$\theta = \begin{cases} -\dfrac{(T - T_C)}{T_{r1} - T_C} & T \leq T_C \\ \dfrac{(T - T_C)}{T_{r2} - T_C} & T \geq T_C \end{cases} \quad (7)$$

where $T_{r1}$ and $T_{r2}$ are two reference temperatures below and above $T_C$ satisfying the relation $\Delta S_M(T_{r1}) = \Delta S_M(T_{r2}) = f \times \Delta S_M^{max}$ with $f = 0.5$ for this study. **Fig. 6b** shows the $\Delta S_M / \Delta S_M^{max}$ vs. $\theta$ curve from 1 – 5 T in which the data indeed collapses onto a universal master curve. It has been suggested that the existence of a universal curve of $\Delta S_M/\Delta S_M^{max}$ vs. $\theta$ is a conclusive proof of the SOMT nature [66]. A universal curve cannot be constructed for the case of FOMT materials.

Finally, to reconfirm the validity of all the critical exponents determined from the MAP and KF methods, these exponents can be tested by the scaling analysis via the static-scaling hypothesis, which relates to the magnetization $M$ and applied magnetic field $\mu_0 H$. The isothermal magnetization data is rescaled based on the renormalized equation of state [67]:

$$m = f_{\pm}(h) \qquad (8)$$

$$h/m = \pm a_{\pm} + b_{\pm} m^2 \qquad (9)$$

where the plus and minus signs depict the temperature ranges above and below $T_C$, respectively; $m \equiv |\varepsilon|^{-\beta} M(H,\varepsilon)$ and $h \equiv |\varepsilon|^{-\beta\delta} H$ are the renormalized magnetization and magnetic field, respectively. The expression $f_{\pm}(h) = M(h, \varepsilon / |\varepsilon| = \pm 1)$ determines two universal curves onto which the rescaled magnetization data should collapse above and below $T_C$ [67]. **Fig. 7a** and **Fig. 7b** shows a good collapse of the rescaled magnetization data based on equation (8) and (9). This collapses confirm all the calculated critical exponents are correct.

## CONCLUSION

In summary, we have studied the structural, magnetic properties, magnetocaloric effect and critical behavior of the polycrystalline $(La_{0.5}Pr_{0.5})_{0.6}Ba_{0.4}MnO_3$. The sample undergoes a second-order paramagnetic to ferromagnetic phase transition around $T_C = 277$ K near room temperature. The maximum magnetic entropy change $-\Delta S_M^{max}$ is ~ 3.22 J kg$^{-1}$ K$^{-1}$ for a field change of 5 T. The enhancement of *RCP* is the result of the large magnetic entropy change and the broadened PM-FM phase transition. A detailed analysis of the critical exponents ($\beta = 0.514 \pm 0.010$, $\gamma = 1.164 \pm 0.013$, and $\delta = 3.265 \pm 0.023$) indicates the long range ferromagnetic interaction below the $T_C$ but the short range interaction above the $T_C$, causing the broadening of the PM-FM phase transition and consequently enhancing the *RCP*.

**Acknowledgments**

The authors acknowledge the financial support of Funds for Science and Technology Development of the University of Danang under Grant Number of B2016-DNA-43-TT. Work at USF was supported by the U.S. Department of Energy, Office of Basic Energy Sciences, Division of Materials Sciences and Engineering under Award No. DE-FG02-07ER 46438 (Magnetocaloric study).

**References**

[1] V. Franco, J.S. Blázquez, B. Ingale, and A. Conde, Annu. Rev. Mater. Res. 42, 305 (2012).

[2] K.A. Gschneidner Jr. and V.K. Pecharsky, Annu. Rev. Mater. Sci. 30, 387 (2000).

[3] V. K. Pecharsky, K. A. Gschneidner, and A.O. Tsokol, Rep. Prog. Phys. 68, 1479 (2005).

[4] M.H. Phan and S.C. Yu, J. Magn. Magn. Mater. 308, 325 (2007).

[5] K.A. Gschneidner Jr and V.K. Pecharsky, International J. Refrigeration 31, 945 (2008).

[6] A. Smith, C.R.H. Bahl, R. Bjork, K. Engelbrecht, K.K. Nielsen, and N. Pryds, Adv. Energy Mater. 2, 1288 (2012).

[7] K.G. Sandeman, Scripta Mater. 67, 566 (2012).

[8] V. Franco, J.S. Blazquez, J.J. Ipus, J.Y. Law, L.M. Moreno-Ramírez, and A. Conde, Prog. Mater. Sci. 93, 112 (2018).

[9] N.S. Bingham, M.H. Phan, H. Srikanth, M.A. Torija, and C. Leighton, J. Appl. Phys. 106, 023909 (2009).

[10] P. Lampen, N. S. Bingham, M.H. Phan, H. Kim, M. Osofsky, A. Piqué, T.L. Phan, S.C. Yu, and H. Srikanth, Appl. Phys. Lett. 102, 062414 (2013).

[11] N.S. Bingham, M.H. Phan, H. Srikanth, M.A. Torija, and C. Leighton, J. Appl. Phys. 126, 023909 (2009).

[12] R.V. Hemolt, J. Wecker, B. Holzapfel, L. Schultz, and K. Samwer, Phys. Rev. Lett. 71, 2231 (1993).

[13] A. Urushibara, Y. Moritomo, A. Asamitsu, G. Kido, and Y. Tokura, Phys. Rev. B. 51, 14103 (1995).

[14] Y. Sun, X.J. Xu, L. Zhang, and Y.H. Zhang, Rev. B. 60, 12317 (1999).

[15] J.C. Debnath, R. Zeng, J.H. Kim, and S.X. Dou, J. Appl. Phys. 107, 09A916 (2010).

[16] Z.U. Rehman, M.S. Anwar, and B.H. Koo, J. Supercond. Nov. Magn. 28, 1629–1634 (2015).

[17] L. Han, A. Zhang, W. Zhai, J. Yang, Z. Yan, and T. Zhang, J. Supercond. Nov. Magn. 32, 10 (2019).

[18] A. Varvescu and I.G. Deac, Physica B: Phys. Condens. Matter 470–471, 96-101 (2015).

[19] A. Barnabé, F. Millange, A. Maignan, M. Hervieu, and B. Raveau, Chem. Mater. 10, 252-259 (1998).

[20] M. Baazaoui, M. Boudard, and S. Zemni, Mater. Lett. 65, 2093 (2011).

[21] D.T. Morelli, A.M. Mance, J.V. Mantese, and A.L. Micheli, J. Appl. Phys. 79, 373 (1996).

[22] W. Zhong, W. Chenb, C.T. Auc, and Y.W. Dua, J. Magn. Magn. Mater. 261, 238 (2003).

[23] I. Hussain, M. S. Anwar, E. Kim, B.H. Koo, and C.G. Lee, Korean J. Mater. Res. Vol. 26, No. 11, 623 (2016).

[24] N. Moutis, I. Panagiotopoulos, M. Pissas, and D. Niarchos, Phys. Rev. B. 59-2, 1129 (1998).

[25] M.H. Phan, V. Franco, N.S. Bingham, H. Srikanth, N.H. Hur, and S.C. Yu, J. Alloy Comp. 508, 238 (2010).

[26] V. Dyakonov, F. Bukhanko, V. Kamenev, E. Zubov, S. Baran, T. Jaworska-Gołąb, A. Szytuła, E. Wawrzyńska, B. Penc, R. Duraj, N. Stüsser, M. Arciszewska, W. Dobrowolski, K. Dyakonov, J. Pientosa, O. Manus, A. Nabialek, P. Aleshkevych, R. Puzniak, A. Wisniewski, R. Zuberek, and H. Szymczak, Phys. Rev. B. 74, 024418 (1996).

[27] J. Philip and T.R.N. Kutty, J. Phys.: Condens. Matter 11, 8537 (1999).

[28] J.J. Urban, L. Ouyang, M.H. Jo, D.S. Wang, and H. Park, Nano Lett. Vol. 4, No. 8, 1547 (2004).

[29] J. Jativa, J.F. Jurado, and C. Vargas-Hernandez, Revista Mexicana de Física S 58-2, 19 (2012).

[30] M.H. Phan and S.C. Yu, J. Magn. Magn. Mater. 308, 325 (2007).

[31] V. Franco, J.S. Blázquez, J.J. Ipus, J.Y. Law, L.M. Moreno-Ramírez, and A. Conde, Prog. Mater. Sci. 93, 112 (2018).

[32] G.H. Jonker and J.H. van Santen, Physica 16-3, 337 (1950).

[33] G.H. Jonker and J.H. van Santen, Physica 22, 707 (1956).

[34] J.H. van Santen and G.H. Jonker, Physica 16-7, 599 (1950).

[35] Y. Tokura and Y. Tomioka, J. Magn. Magn. Mater. 200, 1 (1999).

[36] C. Zener, Phys. Rev. 82, 403 (1951).

[37] P.W. Anderson and H. Hasegawa, Phys. Rev. 100, 675 (1955).

[38] J.A. Millis, B.P. Littlewood, and I.B. Shraiman, Phys. Rev. Lett. 74, 5144 (1995).

[39] E. Dagotto, J. Burgy, and A. Moreo, Solid State Communications 126, 9 (2003).

[40] M.H. Phan, S.B. Tian, S.C. Yu and A.N. Ulyanov, J. Magn. Magn. Mater. 256, 306-310 ( 2003).

[41] M.H. Phan, H.X. Peng, and N.D. Tho, J. Appl. Phys. 99 08Q108, 1-3 (2006).

[42] Y.D. Zhang, P.J. Lampen, T.L. Phan, S.C. Yu, H. Srikanth, and M.H. Phan, J. Appl. Phys. 111, 063918 (2012).

[43] P. Lampen, Y. D. Zhang, T. L. Phan, Y. Y. Song, P. Zhang, S. C. Yu, H. Srikanth, and M.H. Phan, J. Appl. Phys. 112, 113901 (2012).

[44] S. K. Mishra, R. Pandit, and S. Satpathy, Phys. Rev. B 56, 2316 (1997).

[45] M.E. Wood and W.H. Potter, Cryogenics 25, 667 (1985).

[46] H.E. Stanley, Oxford University Press, London (1971).

[47] Z. G. Zheng, X. C. Zhong, Z. W. Liu, D. C. Zeng, V. Franco, and J. L. Zhang, J. Magn. Magn. Mater. 343, 184 (2013).

[48] A. Arrot and J.E. Noakes, Phys. Rev. Lett. 19, 786 (1967).

[49] M. Seeger, S.N. Kaul, H. Kronmuller, and R. Reisser, Phys. Rev. B 51, 12585–12594 (1995).

[50] D. Kim, B. Revaz, B.L. Zink, F. Hellman, J.J. Rhyne, and J.E. Mitchell, Phys. Rev. Lett. 89, 227202 (2002).

[51] B. Widom, J. Chem. Phys. 43, 3898 (1965).

[52] J.S. Kouvel and M.E. Fisher, Phys. Rev. A 136, 1626 (1964).

[53] R. Caballero-Flores, N.S. Bingham, M.H. Phan, M.A. Torija, C. Leighton, V. Franco, A. Conde, T.L. Phan, S.C. Yu, and H. Srikanth, J. Phys.: Condens. Matter 26, 286001 (2014).

[54] T. A. Ho, M. H. Phan, V. D. Lam, T. L. Phan, and S. C. Yu, J. Electr. Mater. 45, 2508 (2016).

[55] A. Biswas, S. Chandra, T. Samanta, B. Ghosh, and M.H. Phan, A. K. Raychaudhuri, I. Das, and H. Srikanth, Phys. Rev. B 87, 134420 (2013).

[56] Y.D. Zhang, P.J. Lampen, T.L. Phan, S.C. Yu, H. Srikanth, and M.H. Phan, J. Appl. Phys. 111, 063918 (2012).

[57] M.H. Phan, H.X. Peng and S.C. Yu, J. Appl. Phys. 97, 10M306 (2005).

[58] M.H. Phan, N.D. Tho, N. Chau, S.C. Yu and M. Kurisu, J. Appl. Phys. 97, 3215-3218 (2005).

[59] M.H. Phan, S.C. Yu and N.H. Hur, Appl. Phys. Lett. 86, 072504-072506 (2005).

[60] R. Caballero-Flores, N.S. Bingham, M.H. Phan, M.A. Torija, C. Leighton, V. Franco, A. Conde, T.L. Phan, S.C. Yu, and H. Srikanth, J. Phys.: Condens. Matter 26, 286001 (2014).

[61] M.H. Phan, V. Franco, A. Chaturvedi, S. Stefanoski, G.S. Nolas, and H. Srikanth, Phys. Rev. B 84, 054436 (2011).

[62] H. Oesterreicher and F. T. Parker, J. Appl. Phys. 55, 4334 (1984).

[63] M. Campostrini, M. Hasenbusch, A. Pelissetto, P. Rossi, and E. Vicari, Phys. Rev. B 65, 144520 (2002).

[64] V. Franco, A. Conde, J. M. Romero-Enrique, and J. S. Blazquez, J. Phys.: Condens. Matter 20, 28520 (2008).

[65] S.K. Banerjee, Phys. Lett. 12, 16-17 (1964).

[66] V. Franco, J.S. Blazquez, and A. Conde, Appl. Phys. Lett. 89, 222512 (2006).

[67] C. M. Bonilla, J.H. Albillos, F. Bartolomé, L.M. García, M.P. Borderías, and V. Franco, Phys. Rev. B 81, 224424 (2010).

**Tables**

**Table I.** Curie temperature $T_C$, applied magnetic field $\mu_0\Delta H$, maximum magnetic entropy change $|\Delta S_M^{max}|$, relative cooling power *RCP* for $(La_{0.5}Pr_{0.5})_{0.6}Ba_{0.4}MnO_3$ and other candidates reported previously in the literature.

| Composition | Structure | $T_C$ (K) | $\mu_0\Delta H$(T) | $-\Delta S_m^{max}$ (J·kg$^{-1}$·K$^{-1}$) | RCP (Jkg$^{-1}$) | Ref |
|---|---|---|---|---|---|---|
| $(La_{0.5}Pr_{0.5})_{0.6}Ba_{0.4}MnO_3$ | Cubic | 276 | 5 | 3.22 | 270 | This work |
| $Pr_{0.8}Ba_{0.2}MnO_3$ | Orthorhombic | 142.5 | 1 | 0.8 | 43.97 | [16] |
| $Pr_{0.7}Ba_{0.3}MnO_3$ | Orthorhombic | 183.3 | 1 | 0.92 | 49.67 | [16] |
| $Pr_{0.6}Ba_{0.4}MnO_3$ | Orthorhombic | 194.87 | 1 | 2.4 | 81.7 | [16] |
| $Pr_{0.5}Ba_{0.5}MnO_3$ | Orthorhombic | 227.9 | 1 | 1.1 | 35.09 | [16] |
| $Pr_{0.87}Ba_{0.13}MnO_3$ | Orthorhombic | 130 | - | - | - | [17] |
| $Pr_{0.77}Ba_{0.23}MnO_3$ | Orthorhombic | 163 | - | - | - | [17] |
| $Pr_{0.67}Ba_{0.33}MnO_3$ | - | 188 | 1 | 2.32 | 49 | [18] |
| | - | 188 | 4 | 5.50 | 225 | [18] |
| $Pr_{0.6}Ba_{0.4}MnO_3$ | - | 235 | - | - | - | [19] |
| $La_{0.6}Ba_{0.4}MnO_3$ | - | 332 | - | - | - | [19] |
| $Pr_{0.67}Ba_{0.33}MnO_3$ | - | 205 | 1 | 1.34 | 28 | [20] |
| | - | 205 | 5 | 4.37 | 230 | [20] |
| $La_{0.67}Ba_{0.33}MnO_3$ | - | 332 | 1 | 0.8 | 40 | [20] |
| | - | 332 | 5 | 3.51 | 235 | [20] |
| $La_{0.67}Ba_{0.33}MnO_3$ | Cubic | 292 | 5 | 1.48 | 161 | [21] |
| $La_{0.67}Ba_{0.33}MnO_3$ | Cubic | 337 | 1 | 2.7 | 68 | [22] |
| $La_{0.7}Ba_{0.3}MnO_3$ | Rhombohedral | 342 | 2.5 | 2.1 | 124 | [23] |
| $La_{0.6}Ba_{0.4}MnO_3$ | Rhombohedral | 333 | 2.5 | 1.192 | 79.31 | [23] |
| $La_{0.67}Ba_{0.33}MnO_3$ | Rhombohedral | 306 | - | - | - | [24] |
| $La_{0.5}Pr_{0.5}MnO_3$ | Cubic | 137 | - | - | - | [26] |
| $La_{0.7}Ba_{0.3}MnO_3$ | Cubic | - | - | - | - | [28] |
| $La_{0.5}Ba_{0.5}MnO_3$ | Cubic | - | - | - | - | [29] |

**Table II.** Curie temperature $T_C$ and critical exponents of $(La_{0.5}Pr_{0.5})_{0.6}Ba_{0.4}MnO_3$ in comparison those of the theoretical models and other materials.

MAP: modified-Arrott plot; KF: Kouvel Fisher; CIA: critical isotherm analysis

| Material/Model | Method | $T_C$(K) | $\beta$ | $\gamma$ | $\delta$ | $n$ | Ref. |
|---|---|---|---|---|---|---|---|
| Mean field | Theory | - | 0.5 | 1.0 | 3.0 | 0.67 | [50] |
| 3D Heisenberg | Theory | - | 0.365 | 1.336 | 4.80 | 0.63 | [50] |
| 3D Ising | Theory | - | 0.325 | 1.241 | 4.82 | 0.57 | [50] |
| Tricritical Mean field | Theory | - | 0.25 | 1.0 | 5.0 | 0.40 | [51] |
| $(La_{0.5}Pr_{0.5})_{0.6}Ba_{0.4}MnO_3$ | MAP | 277 | 0.514 ± 0.010 | 1.164 ± 0.013 | 3.265 ± 0.023 | 0.710 ± 0.033 | This work |
| | KF | 276.56 ± 2.40 | 0.514 ± 0.007 | - | 3.265 ± 0.012 | 0.710 ± 0.019 | This work |
| | KF | 276.68 ± 2.52 | - | 1.164 ± 0.005 | | | This work |
| | CIA | - | - | - | 3.30 ± 0.05 | - | This work |
| | Eq.(6) | 0.561 | 1.240 | - | - | 0.756 ± 0.029 | This work |
| $Pr_{0.87}Ba_{0.13}MnO_3$ | MAP | 130 | 0.459 ± 0.03 | 1.346 ± 0.02 | - | - | [17] |
| | CIA | 130 | - | - | 4.32 ± 0.04 | - | [17] |

| Compound | Method | T (K) | β | γ | δ | | Ref |
|---|---|---|---|---|---|---|---|
| Pr$_{0.77}$Ba$_{0.23}$MnO$_3$ | MAP | 163 | 0.424 ± 0.03 | 1.362 ± 0.02 | - | - | [17] |
| | CIA | 163 | - | - | 4.19 ± 0.03 | | [17] |
| Pr$_{0.67}$Ba$_{0.33}$MnO$_3$ | MAP | 188 | 0.366 | 1.375 | - | - | [18] |
| | CIA | 188 | - | - | 4.743 | - | [18] |
| La$_{0.67}$Ba$_{0.33}$MnO$_3$ | KF | 306 | 0.356 ± 0.004 | 1.120 ± 0.003 | 4.15 ± 0.05 | - | [24] |

**Figure captions**

**Figure 1:** XRD pattern of $(La_{0.5}Pr_{0.5})_{0.6}Ba_{0.4}MnO_3$. Inset shows a SEM image of the sample.

**Figure 2: (a)** Temperature dependences of the magnetization (left hand, in blue color) and magnetic susceptibility (right hand, in red color) at a field of 500 Oe. Inset shows a $dM/dT$ vs. $T$ curve; **(b)** The magnetic field (up to 12 kOe) dependence of magnetization for LPBMO measured at room temperature ($T_{room} = 300$ K).

**Figure 3: (a)** A set of isothermal $M(\mu_0H)$ curves of LPBMO in a temperature range of 100–319 K, with $\delta T_1 = 1$ K from 274 – 285 K and $\delta T_2 = 3$ K for other temperature ranges under applied magnetic fields up to 5 T; **(b)** The temperature dependence of magnetization at various magnetic fields from 0 to 5 T; **(c)** A filled 2-D contour plot of the temperature and applied magnetic field dependence of magnetization; **(d)** The filled 2-D contour plot of the temperature and applied magnetic field dependence of $dM/dT$.

**Figure 4: (a)** The temperature dependence of magnetic entropy change ($\Delta S_M$) for different applied field changes (0.1 - 5 T); **(b)** The applied magnetic field $\mu_o \Delta H$ dependence of the maximum entropy change $-\Delta S_M^{max}$; **(c)** The applied magnetic field $\mu_o \Delta H$ dependence of the full width at half maximum of the $\Delta S_M$ ($T$) curve; **(d)** The applied magnetic field $\mu_o \Delta H$ dependence of the relative cooling power *RCP*.

**Figure 5: (a)** Arrott-Noakes plots using the exponents obtained from the iterative procedure described in the text; **(b)** Spontaneous magnetization $M_S$ and inverse initial susceptibility $\chi_0^{-1}$ as a function of reduced temperature $\varepsilon$, determined from extrapolation of the Arrott-Noakes plots and **(c)** Kouvel-Fisher plots of magnetization data. Straight lines are linear fits to the data, from which $\beta$, $\gamma$, $T_{C1}$, and $T_{C2}$ were computed. The final value of $T_C$ is taken as the average of $T_{C1}$ and $T_{C2}$; **(d)** $\ln(M)$ vs $\ln(\mu_0 H)$ for temperatures near the critical isotherm.

**Figure 6: (a)** The maximum magnetic entropy change $\Delta S_M^{max}$ vs. $(\mu_0 \Delta H)^n$, where $n = 0.756$ is the prediction of the scaling relation; **(b)** A universal $\Delta S_M / \Delta S_M^{max}$ vs. $\theta$ curve.

**Figure 7:** Rescaled magnetization isotherms according to equations of state given in **(a)** equation (8) and **(b)** equation (9).

**Figure 1**

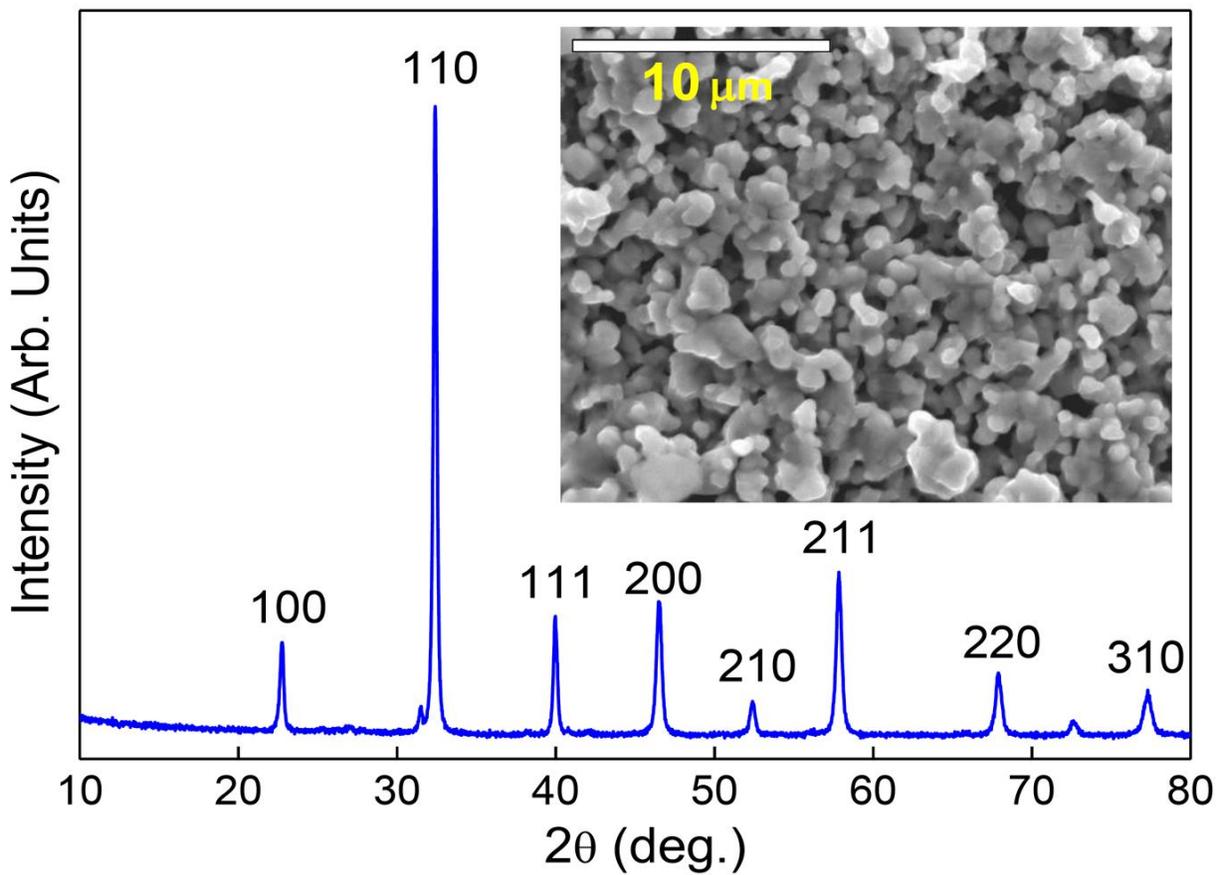

**Figure 2**

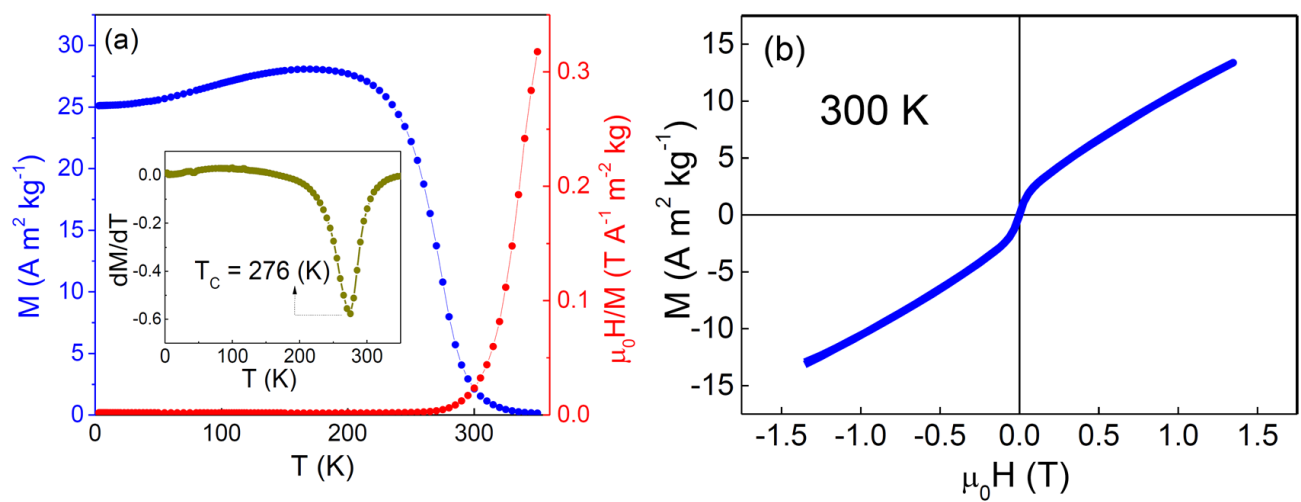

**Figure 3**

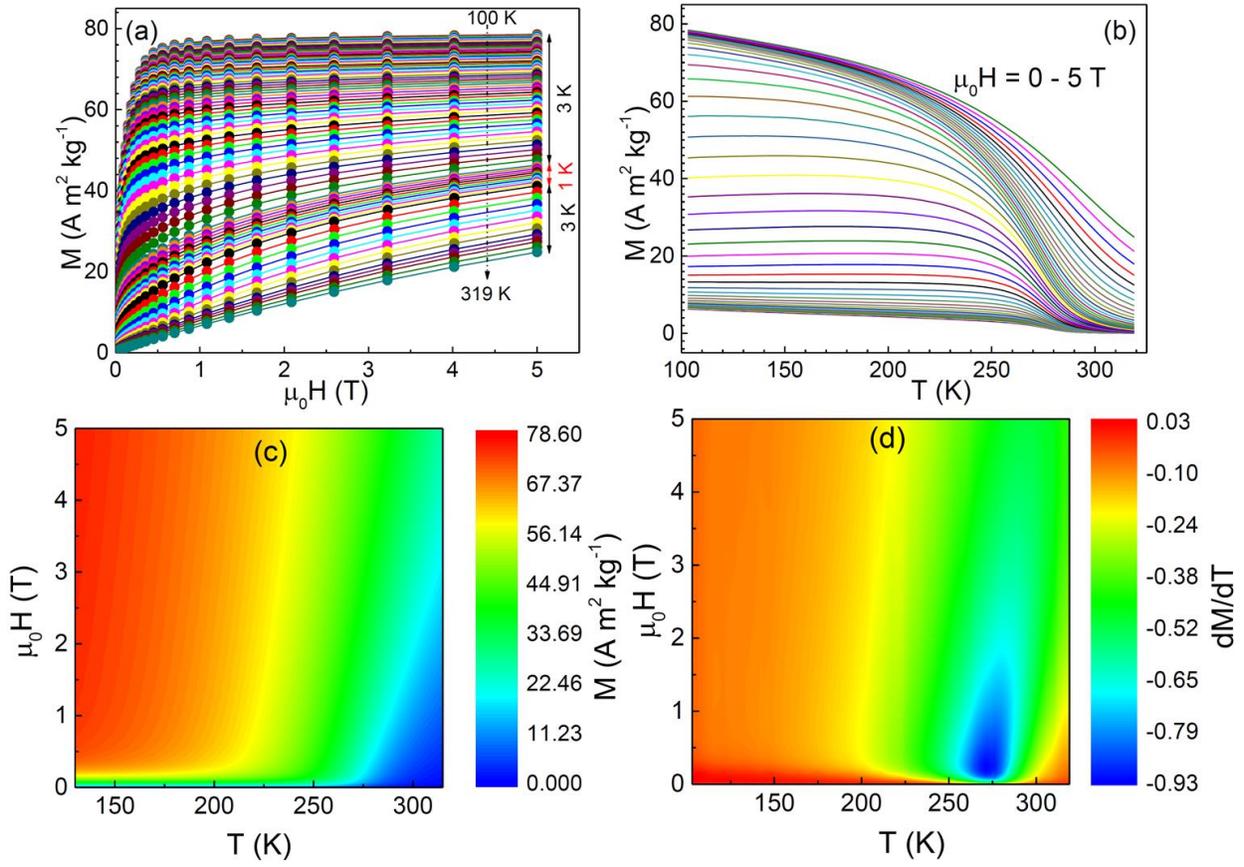

**Figure 4**

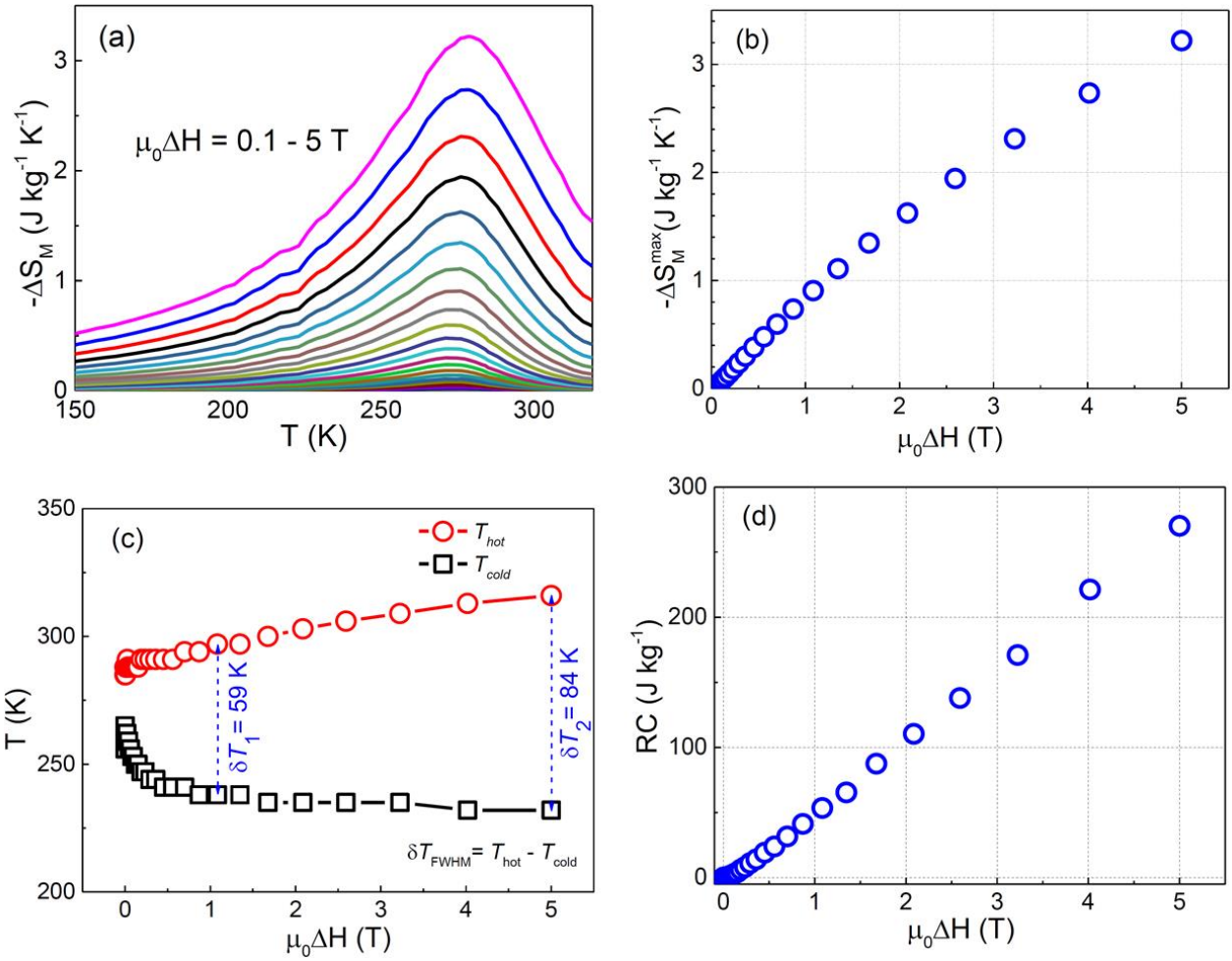

**Figure 5**

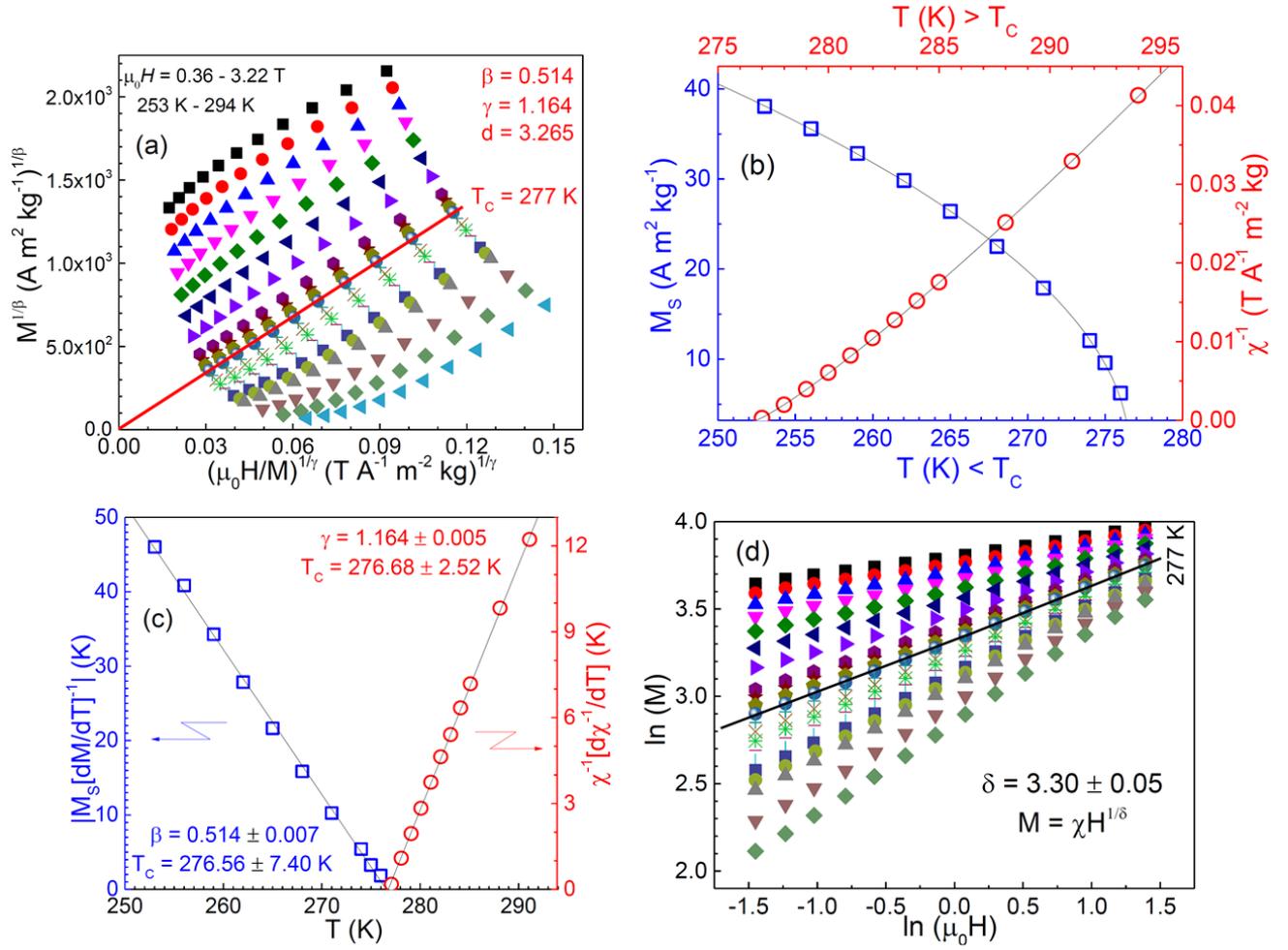

**Figure 6**

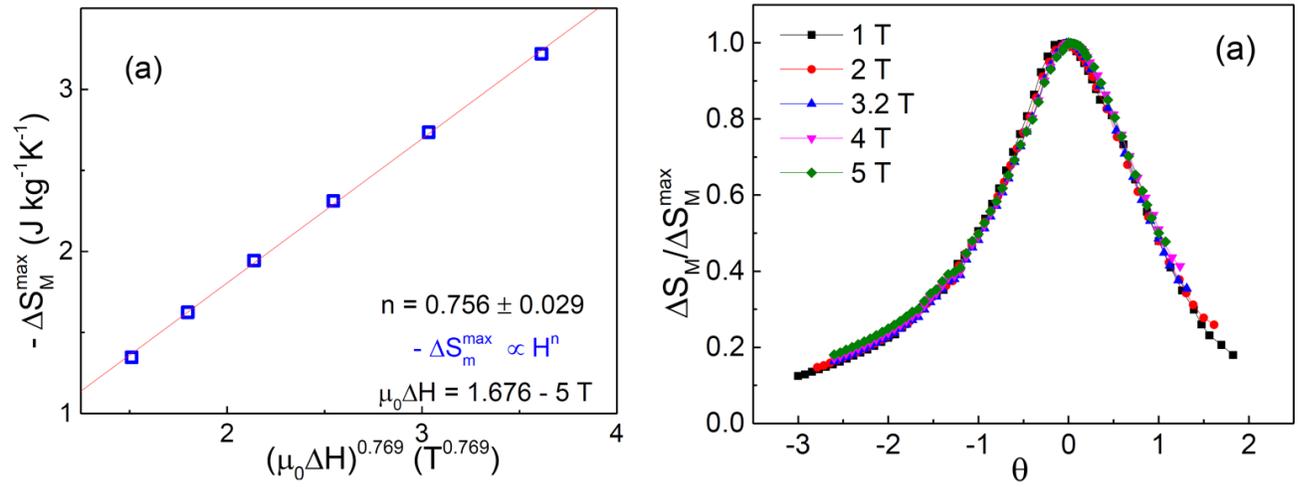

**Figure 7**

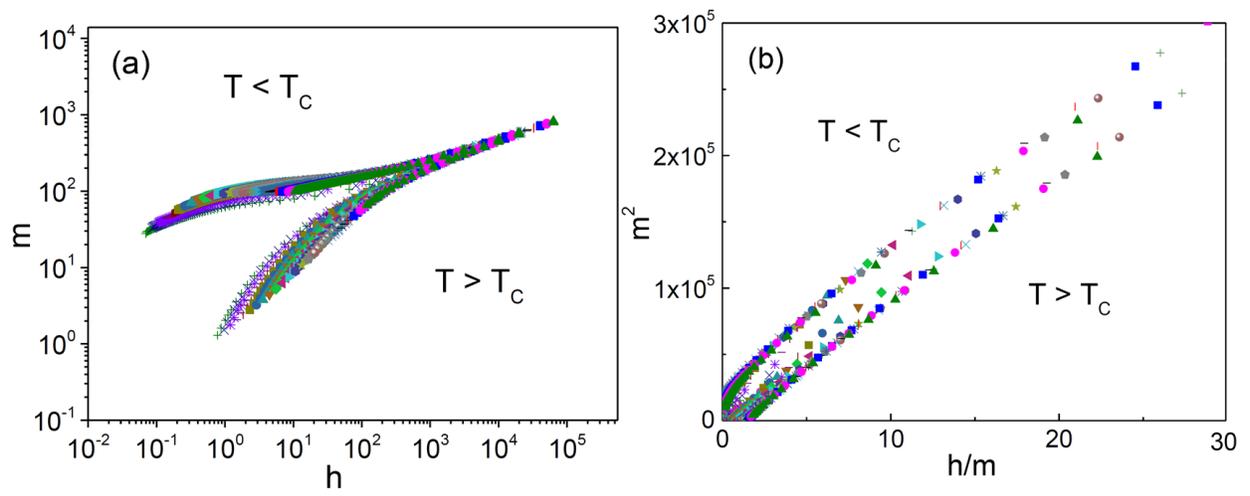